\documentclass[11pt,preprint]{aastex}

\def\ltsima{$\; \buildrel < \over \sim \;$}
\def\lsim{\lower.5ex\hbox{\ltsima}}
\def\gtsima{$\; \buildrel > \over \sim \;$}
\def\gsim{\lower.5ex\hbox{\gtsima}}

\slugcomment{}

\shorttitle{}
\shortauthors{Winn et al.}

\begin{document}

\title{KH~15D:\ Gradual Occultation of a Pre--Main-Sequence Binary}

\author{
Joshua N.\ Winn\altaffilmark{1,2},
Matthew J.\ Holman\altaffilmark{1},
John A.\ Johnson\altaffilmark{3},\\
Krzysztof Z.\ Stanek\altaffilmark{1},
Peter M.\ Garnavich\altaffilmark{4}
}

\altaffiltext{1}{Harvard-Smithsonian Center for Astrophysics, 60
Garden St., Cambridge, MA 02138}

\altaffiltext{2}{National Science Foundation Astronomy \& Astrophysics
Postdoctoral Fellow}

\altaffiltext{3}{Department of Astronomy, University of California,
Berkeley, CA 94720}

\altaffiltext{4}{University of Notre Dame, Notre Dame, IN 46556}

\begin{abstract}
We propose that the extraordinary ``winking star'' KH~15D is an
eccentric pre--main-sequence binary that is gradually being occulted
by an opaque screen. This model accounts for the periodicity, depth,
duration, and rate of growth of the modern eclipses; the historical
light curve from photographic plates; and the existing radial velocity
measurements. It also explains the re-brightening events that were
previously observed during eclipses, and the subsequent disappearance
of these events. We predict the future evolution of the system and its
full radial velocity curve. Given the small velocity of the occulting
screen relative to the center of mass of the binary, the screen is
probably associated with the binary, and may be the edge of a
precessing circumbinary disk.
\end{abstract}


\section{Introduction}
\label{sec:intro}

While monitoring stars in the young cluster NGC~2264, Kearns \& Herbst
(1998) noticed that star \#15 in field D of their sample had a bizarre
light curve. This object, known as KH~15D, undergoes periodic eclipses
($P=48.35$~days) that are remarkable for their depth (3.5~mag) and
duration (currently $\approx$24~days). There is consensus that the
eclipses are caused by circumstellar material, but not on the
composition or spatial distribution of that material. Theories include
an edge-on circumstellar disk (Hamilton et al.\ 2001, Herbst et al.\
2002, Agol et al.\ 2003, Winn et al.\ 2003), an orbiting vortex of
solid particles (Barge \& Viton 2003), and an asymmetric common
envelope (Grinin \& Tambovtseva 2002). The system has attracted the
attention of numerous observers because the fortuitous alignment may
allow unique studies of circumstellar (or even protoplanetary)
processes, and because the occulting edge can be used as a ``natural
coronagraph'' to map out the environment of the underlying T~Tauri
star (Hamilton et al.\ 2003; Agol et al.\ 2003; Deming, Charbonneau,
\& Harrington 2003).

In this {\it Letter} we present the first quantitative model that
accounts for all the observed properties of the system. Our main
inspiration was the discovery by Johnson \& Winn (2004) that in 1970,
the eclipses appeared to be diluted by the light of a second
star. Building on the demonstration by Herbst et al.\ (2002) that the
ingress and egress light curves can be reproduced by a knife edge
crossing the face of a star, we show that the entire archival and
modern light curves can be reproduced by a knife edge crossing the
orbit of a pair of pre--main-sequence stars.

In \S~2 we review the peculiar phenomenology of KH~15D and describe
how it emerges naturally from the model. In \S~3 we determine
quantitative fits to the data. Finally, in \S~4 we discuss the
physical interpretation of the model and predict the results of future
investigations into this intriguing system.

\section{Qualitative description of the model}
\label{sec:idea}

We would like to understand the following characteristics of KH~15D.
Every $48.35\pm 0.02$ days, it decreases in brightness from $I=14.47$
to $\approx$18 (Hamilton et al.\ 2001 and Herbst et al.\ 2002,
hereafter H01 and H02). Currently, the faint state (``eclipse'') lasts
for approximately half of the photometric period, and the eclipse
duration is increasing by about 1 day~year$^{-1}$ (H02). During
eclipses, the flux has been observed to rise and fall abruptly. In
1995-96, these ``re-brightenings'' briefly returned the system to its
uneclipsed flux, or even brighter. Since 1997, the maximum flux during
these events has decreased monotonically (H01, H02).

Archival observations from 1913 to 1950 show that the system spent
$\lesssim$20\% of the time $>$1 mag fainter than its modern bright
state (Winn et al.\ 2003). Between 1967 and 1982, the system
alternated from bright to faint with the same period as observed
today, but the fractional variation was smaller ($\Delta I=0.67\pm
0.07$) and the bright state was $0.90\pm 0.15$~magnitude brighter
(Johnson \& Winn 2004, hereafter JW04). There appears to be a phase
shift of $\approx$180$\arcdeg$ between the 1967-82 light curve and the
modern light curve, i.e., the modern bright states have nearly the
same phase as the previous faint states (JW04).

All these properties are straightforward consequences of the following
model. Consider two stars, A and B, with a projected orbit depicted in
Figure~1. An opaque screen with a sharp edge oriented vertically
(along the $y$-axis) gradually covers the system, moving from left to
right (increasing $x$). Whenever the orbital motion of a star carries
it to the left of the edge ($x_{\rm star} < x_{\rm screen}$), the
starlight is blocked.

\begin{figure}
\epsscale{1.0}
\plotone{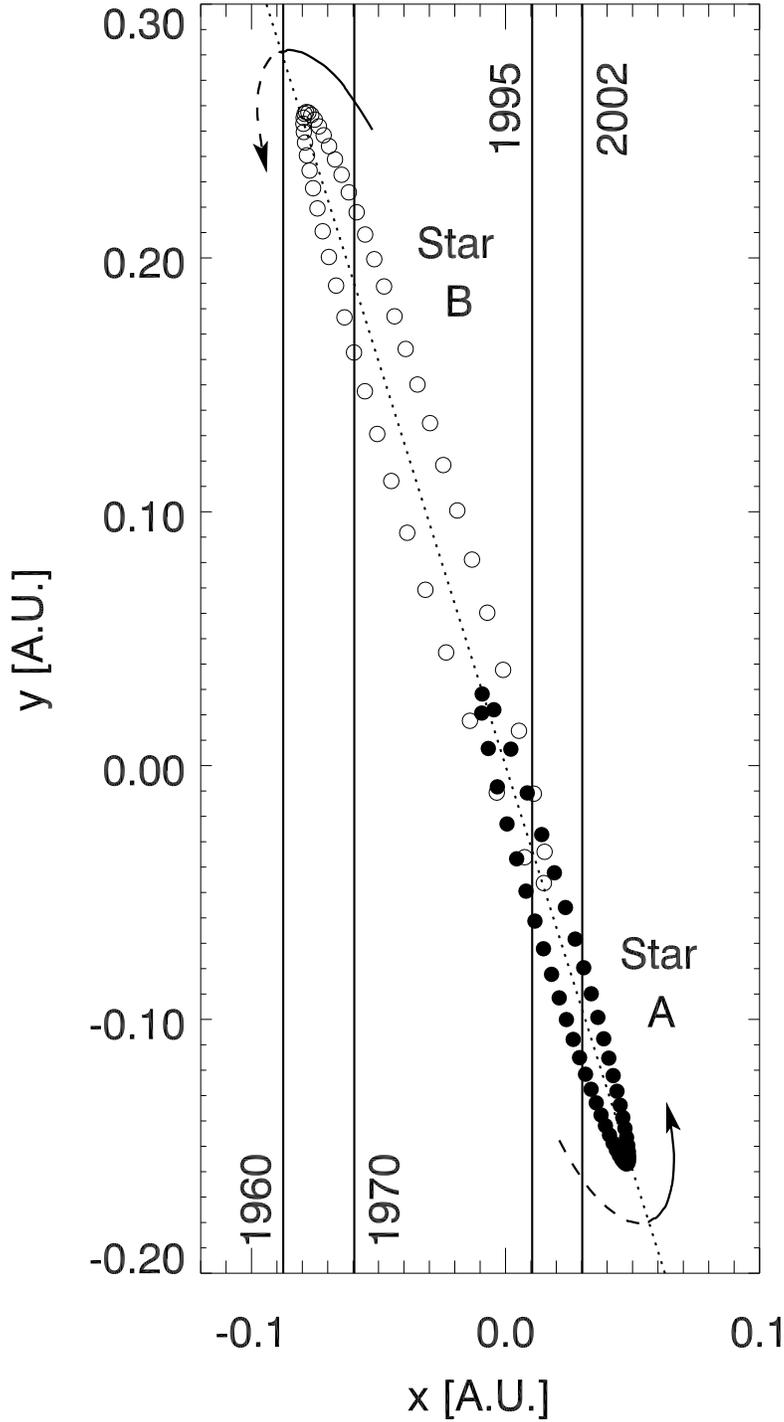}
\caption{ Projected binary orbit of Model 1. The center of mass
defines the origin, and the $z$-axis points towards the observer. The
positions of stars A and B are plotted with a time sampling
$P/50$. The dashed line is the line of nodes, and the arrows show the
motion of the stars between $z>0$ (solid) and $z<0$ (dashed). Vertical
lines indicate the edge of the obscuring screen at some years of
interest. The screen moves from left to right, blocking the stars
whenever they are to the left of the edge. }
\label{fig1}
\end{figure}

Before 1960, both stellar orbits were fully exposed, and no eclipses
were seen, in compliance with the archival observations. In 1970, the
screen covered the left end of B's orbit, but A remained
unobscured. As a result, diluted eclipses of star B were observed. By
2002, the screen covered B's entire orbit and a significant fraction
of A's orbit, causing today's periodic, long-lasting, total
eclipses. The eclipses grow in duration as the screen continues to
advance. The re-brightenings occurred during the time span around
1995, when B's orbit had not yet been completely covered, allowing B
to peek out briefly from behind the screen while A was
eclipsed. Because the eclipses in 1970 were of star B, whereas the
modern eclipses are of star A, this model produces the phase shift
that was tentatively identified by JW04.

\section{Quantitative determination of parameters}
\label{sec:params}

Rather than fitting the model to the voluminous photometric data, we
attempted to match some key derived quantities: the eclipse duration
in 1967-70, the eclipse duration and its rate of increase from 1997 to
2003, and the ingress and egress durations of 2002 (see Table~1). We
also fitted the radial velocity difference of $3.3\pm 0.6$~km~s$^{-1}$
that Hamilton et al.\ (2003) measured near the start and end of a
particular eclipse in 2001.

We assumed the orbital period is $P=48.35$~days and star A has mass
$0.6M_{\odot}$ and radius $1.3R_{\odot}$ (H01). The orbit was
specified by the mass ratio, eccentricity, and sequential rotations
$\theta_z$, $\theta_y$, and $\theta_x$ about the Cartesian axes
defined in Figure~1. By using Cartesian axes, rather than traditional
orbital elements, we could capitalize on certain degeneracies of the
model. Rotation about the $x$-axis does not affect the light
curve. Rotation about the $y$-axis affects the projected size of the
orbits, but not the projected size of the (spherical) stellar
surfaces. Thus, $\theta_y$ controls the time scale of changes in
eclipse duration relative to the time scales of partial-eclipse
phenomena: ingress, egress, and the suppression of re-brightening
events.

For a given orbit, we produced a model light curve as follows. The
speed of the occulting edge relative to the center of mass was
determined by requiring the edge to cross B's orbit between 1963 and
1997. (The actual starting year of the eclipses is unknown.) We
computed the positions of the screen and stars with a time sampling of
$P/100$. We assumed $L_B/L_A=1.3$ (the best fit to the JW04 light
curve) and added a time-independent flux of $0.04 L_A$ to represent
scattered light. Whenever $x_{\rm star} < x_{\rm screen} - R_{\rm
star}$, that star made no contribution to the light curve. When the
screen covered only part of a stellar surface, we used the same
limb-darkening law as H02 ($\mu=0.3$) to compute the flux from that
star.

First, we optimized $M_B$, $e$ and $\theta_z$, by fitting the eclipse
durations and the ratio of durations of ingress and egress. We
searched a 3-d grid of models with $0.5<M_B/M_A<2.0$, $0.05<e<0.95$,
and $-90\arcdeg < \theta_z < 90\arcdeg$. The best-fitting solution had
$\chi^2/N_{\rm D.O.F.} = 0.7$. Next, we adjusted $\theta_y$ to
increase the ingress and egress durations to the observed values. Due
to symmetry, there were two solutions, differing only in sign. We
refer to these as Model 1 and Model 2. For each model, we forced
agreement with the radial velocity measurements by tuning $\theta_x$
and the heliocentric radial velocity of the center of mass. Finally,
the radius of star B was tailored to match the ingress and egress
durations in 1967-70, and the time scale over which the re-brightening
events decreased in intensity.

The parameters of Models 1 and 2, after translating into traditional
orbital elements, are given in Table~2. These models produce identical
light curves, but have different three-dimensional orbits and radial
velocity curves. (Two additional solutions are obtained by reflecting
Models 1 and 2 in the $xz$-plane.) The orbit of Model 1 is depicted in
Figure~1. Figure~2 shows the time evolution of eclipse durations and
ingress/egress ratio. Figure~3 compares the model to the archival
light curves. The agreement is excellent within the overall
uncertainties in zero point and phase of the archival data. Figure~4
shows the disappearance of the re-brightenings between 1995 and 1998,
and compares the model with the 2001-02 light curve. Figure~5 shows
the two possible solutions for the radial velocity curve.

\section{Discussion}
\label{sec:discussion}

We have shown it is possible to find reasonable stellar and orbital
parameters that bring the model into quantitative agreement with the
modern and archival light curves, and the available radial velocity
data. Certainly we do not claim that our model parameters are exactly
correct. The fitted parameters are subject to our assumptions about
the mass and radius of star A, the year the eclipses began, and the
constancy of the screen speed, among other things. Nevertheless, we
can use the generic properties of the model to make some inferences
and predictions about the system.

The speed of the screen relative to the center of mass of the binary
is only $v_x = 13$~m~s$^{-1}$, a scale set by the $\approx$35 year
crossing time of star B's projected orbit. This speed seems too small
for the screen to be an unrelated foreground cloud, as has been
proposed for the central object of planetary nebula NGC~2346 (Schaefer
1985, Mendez et al.\ 1985, Roth et al.\ 1984). Rather, the screen is
probably physically associated with the binary, as also suggested by
the small angle between the screen's edge and the orbital plane of the
binary. A circular orbit with speed 13~m~s$^{-1}$ around a total mass
of 1.0$M_{\odot}$ would have a radius of 25 parsecs, which is too
large to remain bound to the system. The speed must represent a
phenomenon slower than orbital motion, such as orbital precession.

Hence our hypothesis is that the screen is a precessing circumbinary
disk. As a feasibility test, we numerically integrated the motion of
test particles in a circular, circumbinary orbit around the binary
system of Model 1. For orbits of radius 2.6 A.U.\ that are inclined by
$\approx$20\arcdeg\ relative to the plane of the binary, the line of
nodes regresses at $6.3\times 10^{-3}$~rad~year$^{-1}$, corresponding
to a projected velocity of 14~m~s$^{-1}$ normal to the plane of the
binary. Furthermore, given the mass ratio and eccentricity of the
binary, test particles at 2.6~A.U.\ are long-term dynamically stable
against ejection (Holman \& Wiegert 1999).

One difference between the model and the 2001-02 light curve is that
the start of ingress, and end of egress, are too abrupt in the
model. This is probably due to the assumption of a perfectly sharp and
straight edge. Likewise, we did not attempt to model temporal
variations in the scattered light during eclipses, which could be due
to the spatial distribution of dust in the plane of the binary, or
ahead of the occulting edge.

Radial velocity predictions are given in Figure~5, subject to the
uncertainty in the total stellar mass, and possible contamination of
the H02 measurements by scattered light. Soderblom et al.\ (1999)
found a median radial velocity of $+21.2\pm 1.8$~km~s$^{-1}$ for 18
probable members of NGC~2264, favoring Model 1 ($+15.5$~km~s$^{-1}$)
over Model 2 ($+5.7$~km~s$^{-1}$). In either case, the peak-to-peak
variations outside of eclipses should be $\approx$10~km~s$^{-1}$.
Larger variations occur during eclipses, when the star's photosphere
is hidden.

We can also predict the results of future archival studies and
monitoring campaigns. Prior to about 1960, no photometric variations
were seen, apart from any intrinsic variations of the stars. The
eclipse duration grew between 1970 and 1985, at which point the light
curve became more complex due to eclipses of both stars. In the future
the eclipse duration will continue to grow, and by about 2012 the
system will be completely covered. When the system will come back into
view depends on the unknown extent of the obscuring screen.

\acknowledgments We are very grateful to Cesare Barbieri, Scott Gaudi,
Catrina Hamilton, Bill Herbst, Marc Kuchner, Lucas Laursen, Francesca
Rampazzi, Dimitar Sasselov, Brad Schaefer, and Martin Winn, for
helpful discussions and assistance. This work was supported by the
National Science Foundation under Grant No.\ 0104347.

\begin{figure}[h]
\epsscale{0.9}
\plotone{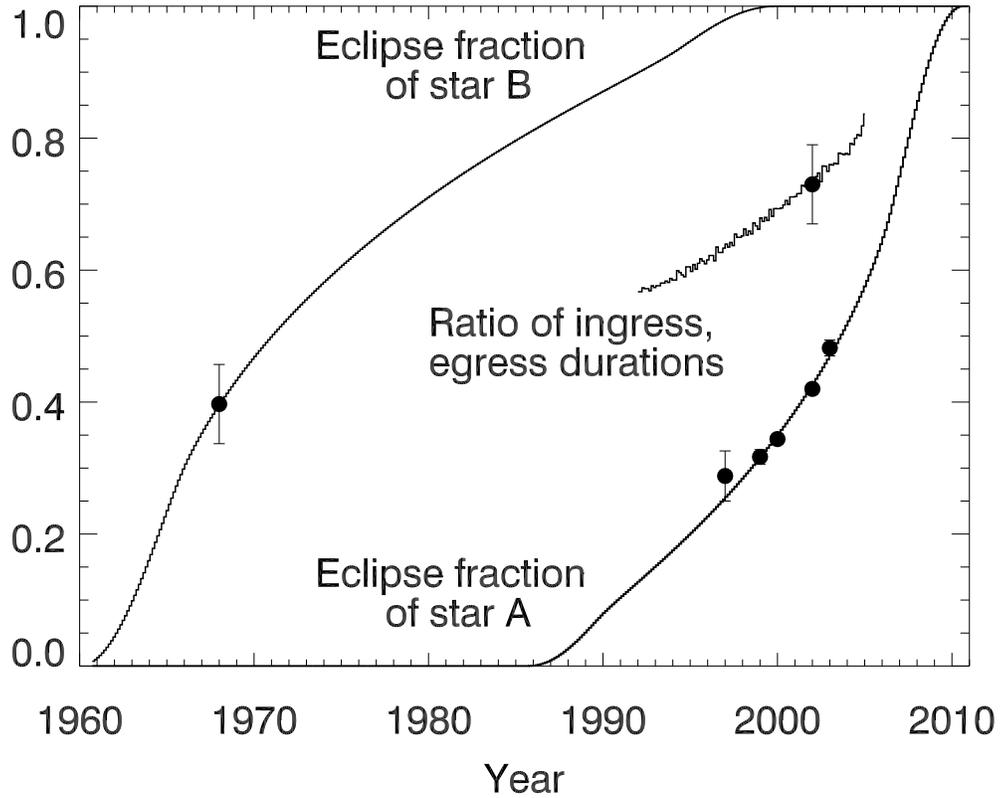}
\caption{ Time evolution of the eclipse fractions of stars A and B,
and the ratio of ingress and egress durations of star A, for the
best-fit model. The solid circles are observed values (see Table~1). }
\label{fig2}
\end{figure}

\begin{figure}
\epsscale{0.8}
\plotone{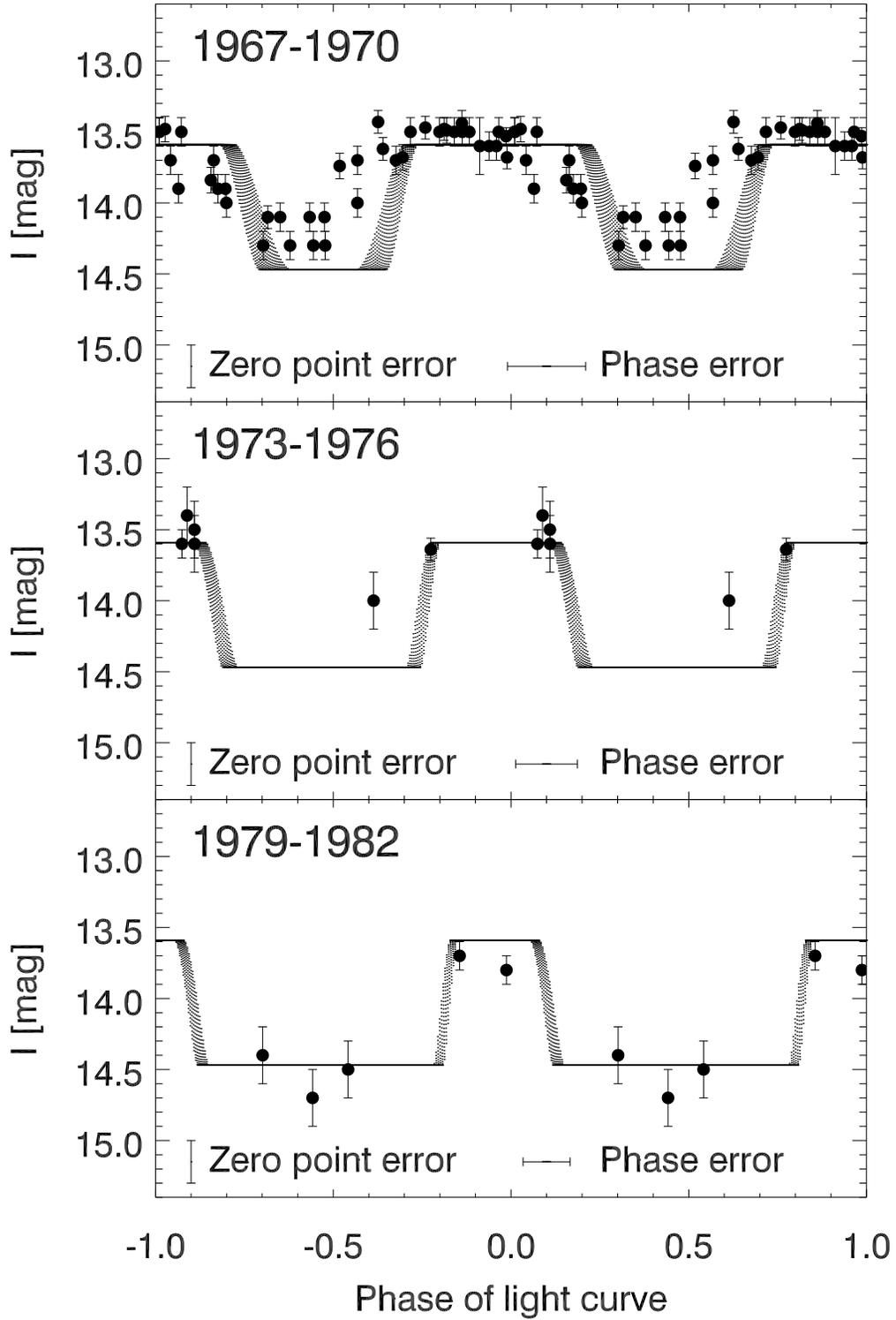}
\vspace{0.2in}
\caption{ Comparison of phased light curves from the model (dotted
lines) and data (solid circles) from the Asiago Observatory archive
(Barbieri et al.\ 2003). The light curves were phased with the H02
ephemeris. The thickness of the model light curves during ingress and
egress is due to the growth in eclipse duration during the 3-year time
spans. }
\label{fig3}
\end{figure}

\begin{figure}
\epsscale{0.75}
\plotone{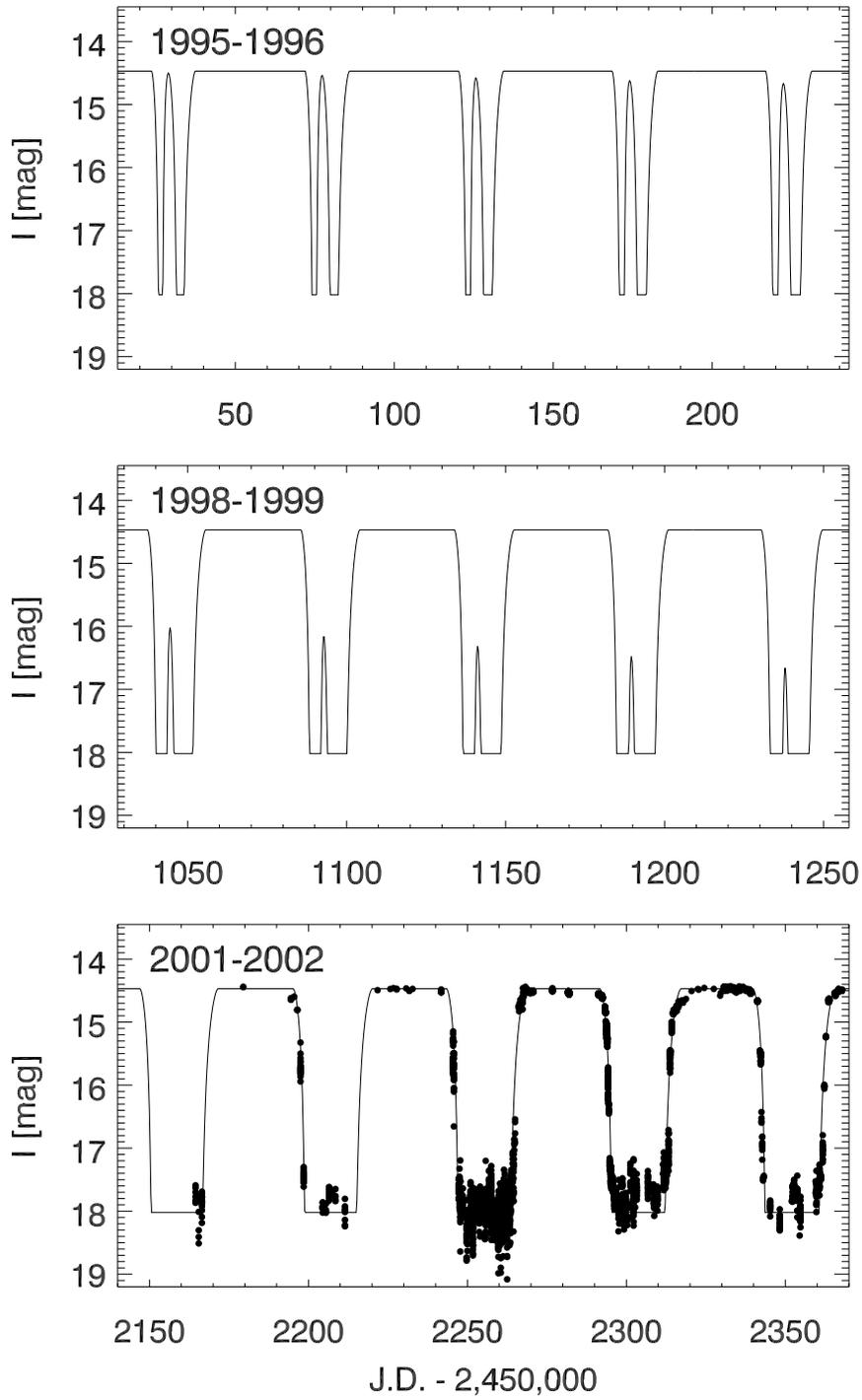}
\vspace{0.3in}
\caption{ Model light curves from three recent time periods. The top
and middle panels show the gradual diminution of the amplitude of the
re-brightening events. The bottom panel compares the model and the H02
data. }
\label{fig4}
\end{figure}

\begin{figure}
\epsscale{1.0}
\plotone{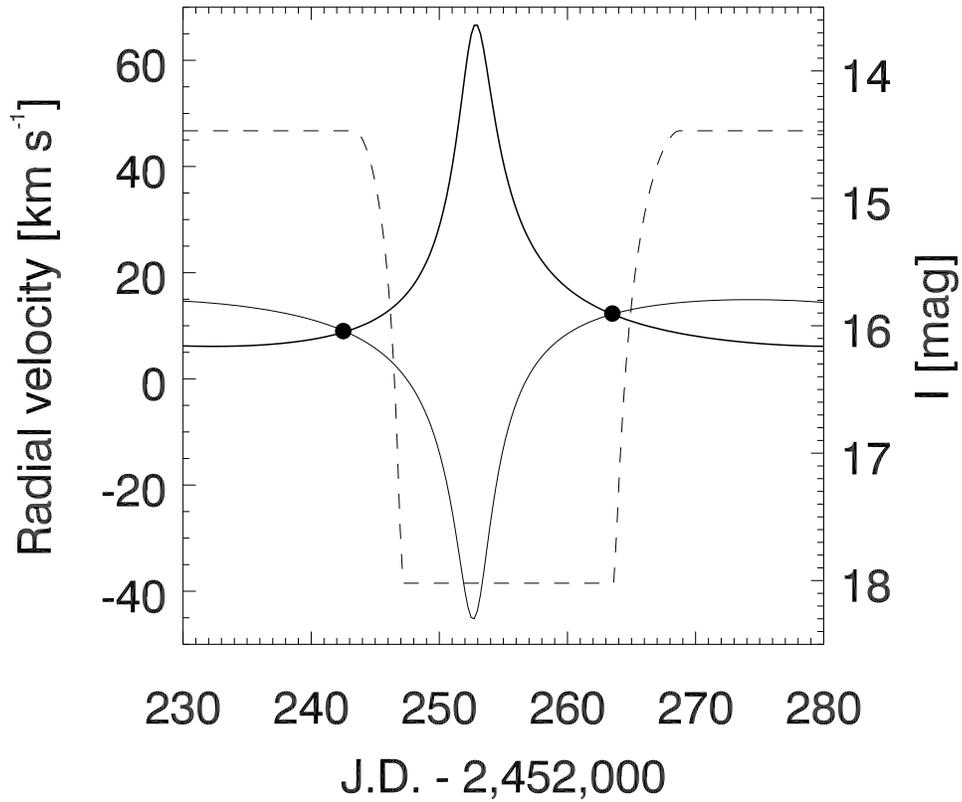}
\caption{ Two possibilities for the radial velocity variation of star
A (solid lines, left-hand axis). Positive velocity corresponds to
motion away from the Sun.  Model 1 corresponds to the positive radial
velocity peak, and Model 2 corresponds to the negative peak. Also
plotted is the corresponding $I$ magnitude of the model light curve
(dashed line, right-hand axis). Solid circles are measurements by
Hamilton et el.\ (2003). }
\label{fig5}
\end{figure}

\begin{deluxetable}{lccc}
\tabletypesize{\small}
\tablecaption{Model constraints}
\tablewidth{0pt}
\tablehead{
\colhead{Quantity} &
\colhead{Star} &
\colhead{Year} &
\colhead{Value\tablenotemark{a}}
}
\startdata
Eclipse fraction\tablenotemark{b} & B & 1968.0 & $0.397\pm 0.060$ \\
Eclipse fraction\tablenotemark{b} & A & 1997.0 & $0.288\pm 0.038$ \\
Eclipse fraction\tablenotemark{b} & A & 1999.0 & $0.317\pm 0.011$ \\
Eclipse fraction\tablenotemark{b} & A & 2000.0 & $0.344\pm 0.008$ \\
Eclipse fraction\tablenotemark{b} & A & 2002.0 & $0.420\pm 0.008$ \\
Eclipse fraction\tablenotemark{b} & A & 2003.0 & $0.482\pm 0.012$ \\
Ingress duration\tablenotemark{c} & A & 2002.0 & $3.8\pm 0.3$~days \\
Egress duration\tablenotemark{c}  & A & 2002.0 & $5.2\pm 0.3$~days \\
\enddata
\tablenotetext{a}{ All quantities and their uncertainties were
estimated with pencil and ruler from the light curves of H02, JW04,
and P.\ Garnavich (unpublished). }
\tablenotetext{b}{ Eclipse fraction is the fraction of the photometric
period during which the total flux is below 52\% of the maximum. }
\tablenotetext{c}{ Ingress and egress durations were estimated by
measuring the time between the $I=14.6$ and 17.6 levels, and then
multiplying by 1.33, to match our choice of limb-darkening model. }
\end{deluxetable}

\begin{deluxetable}{lcc}
\tabletypesize{\small}
\tablecaption{Optimized model parameters}
\tablewidth{0pt}
\tablehead{
\colhead{Parameter} &
\colhead{Model 1} &
\colhead{Model 2}
}
\startdata
Mass ratio, $M_B/M_A$                        & 0.61             & 0.61          \\
Radius of star B [solar radii]               & 1.8              & 1.8           \\
Eccentricity, $e$                            & 0.70             & 0.70          \\
Longitude of ascending node\tablenotemark{b}, $\Omega$ [deg]
                                             & $-72\fdg6$       & $105\fdg9$    \\
Argument of pericenter, $\omega$ [deg]       & $-7\fdg2$        & $184\fdg9$    \\
Inclination, $i$ [deg]                       & $84\fdg6$        & $81\fdg0$     \\
Heliocentric radial velocity of C.O.M.\tablenotemark{a}\,\, [km~s$^{-1}$]
                                             & $+15.5$          & $+5.7$ \\
Relative velocity of screen and C.O.M.\ [m~s$^{-1}$]
                                             & 13               & 13
\enddata
\tablenotetext{a}{ Positive velocity corresponds to motion away from
the Sun. }
\tablenotetext{b}{ Measured counter-clockwise, relative to the
$x$-axis, as in Figure~1. The orientation of the occulting edge in
celestial coordinates is unknown. }
\end{deluxetable}

\end{document}